\documentclass[]{spie}  %>>> use this instead for A4 paper

\usepackage{amsmath,amsfonts,amssymb}
\usepackage{booktabs}
\usepackage{graphicx}
\usepackage{multirow}
\usepackage{array}
\usepackage[colorlinks=true, allcolors=blue]{hyperref}

\title{Hierarchical Visual Interface for Educational Video Retrieval and Summarization}

\author[a]{Jiahao Weng}
\author[b]{Chao Zhang}
\author[c]{Xi Yang}
\author[a]{Haoran Xie}
\affil[a]{Japan Advanced Institute of Science and Technology, Ishikawa, Japan}
\affil[b]{University of Fukui, Fukui, Japan}
\affil[c]{Jilin University, Changchun, China}

\authorinfo{Further author information: (Send correspondence to H. Xie)\\J.Weng: E-mail: s2110022@jaist.ac.jp\\  C. Zhang: E-mail: zhang@u-fukui.ac.jp\\  X. Yang: E-mail: earthyangxi@gmail.com\\  H. Xie: E-mail: xie@jaist.ac.jp, Telephone: +81-0761-51-1753}

\begin{document} 
\maketitle

\begin{abstract}
With the emergence of large-scale open online courses and online academic conferences, it has become increasingly feasible and convenient to access online educational resources. However, it is time consuming and challenging to effectively retrieve and present numerous lecture videos for common users. In this work, we propose a hierarchical visual interface for retrieving and summarizing lecture videos. Users can utilize the proposed interface to effectively explore the required video information through the results of the video summary generation in different layers. We retrieve the input keywords with the corresponding video layer with timestamps, a frame layer with slides, and the poster layer with summarization of the lecture videos. We verified the proposed interface with our user study by comparing it with other conventional interfaces. The results from our user study confirmed that the proposed interface can achieve high retrieval accuracy and good user experience.
\end{abstract}


\section{INTRODUCTION}
\label{sec:intro}  % \label{} allows reference to this section

Open online courses and online academic conferences play an important role in addressing the global pandemic issue. There is a growing trend of courses and academic conferences being held online with available online video sources. However, a major challenge with the emergence of large-scale online resources is how to effectively retrieve and present the desired video contents from numerous resources for end users. While the conventional interface can help navigate educational videos, provide content-based lecture video retrieval, and use data-driven interaction techniques for improving the navigation of Educational Videos~\cite{yadav2016vizig,kim2014data}, it is difficult for the common user to explore videos with satisfied interaction and summarization. For example, when users want to search for a specific technique in an academic conference, they will likely get several results after typing a keyword. The current video retrieval systems do not provide a satisfactory interaction and summarization function for saving the user's time.

 \begin{figure} [ht]
   \begin{center}
   \begin{tabular}{c}
   \includegraphics[width=0.85\textwidth]{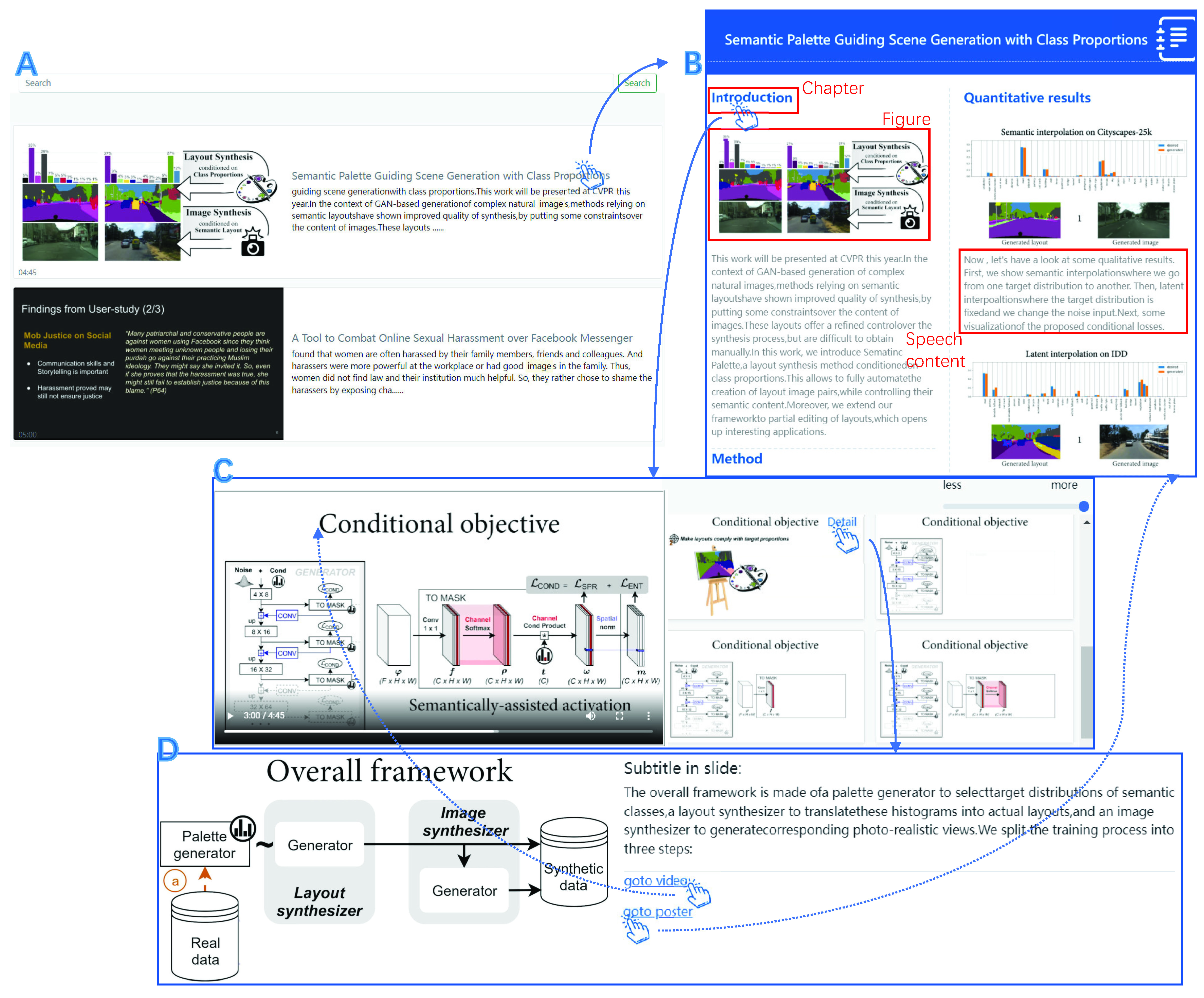}
   \end{tabular}
   \end{center}
   \caption[example] 
   { \label{fig:ui} 
The proposed hierarchical visual interface for lecture videos with retrieval page with keywords (A), poster layer (B), video layer (C) and frame layer (D). }
 \end{figure} 
 
In this paper, we propose a hierarchical visual interface for lecture videos and academic presentations with video retrieval and summarization as shown in Figure \ref{fig:ui}. The proposed system utilizes the slide detection algorithm to extract slides from the video and adopts state-of-the-art image analysis approach \cite{shen2021layoutparser} to extract visual contents from the slides. The hierarchical visual interface consists of a video layer that contains video and groups of slides with timestamps, a frame layer that contains the speech content and visual content of one selected slide, and a poster layer that demonstrates the summarized video in a poster style. This interface can fulfill the diverse usage scenario for different users. The first scenario is that when users want to see an overview of a specific video, they can enter the poster layer.The second scenario is that when users only want to view a specific part of the video, they can click the title in the poster layer to redirect them to the corresponding video layer.The third scenario is that when users want to view the textual matter of the speech content or the slide rather than the video, they can enter the frame layer. Finally, we verify that the proposed visual interface can help the user retrieve lecture videos more conveniently and efficiently.

\section{RELATED WORK}
Given that academic conference presentations and lectures are increasingly being made available on YouTube and Massive Open Online Course (MOOC) websites, the video retrieval from video contents has become particularly common and plays an important role in personal study. These websites obtain metadata from the video and create an algorithm for estimating places on the slide where the speaker explains automatically \cite{tsujimura17_interspeech}. DynamicSlide designed an interaction interface that uses the speech content in the lecture video to show where the lecturer explains on the slide and helps the user take notes \cite{jung2018dynamicslide}. Although these user interfaces may help users in online lecture learning, they may have difficulty with retrieving the information conveniently.

A data-driven interaction technique was proposed for navigating educational videos \cite{kim2014data}. The proposed method can improve the current video user interface by introducing a timeline with interaction peaks to show points of  high user activity and personal watching traces. A novel voice navigation for how-to videos was also proposed \cite{chang2021rubyslippers} to allow users to search for content by saying a keyword. The system provides user recommendation suggestions for each interaction scenario. With the interaction technique, users can directly obtain the result through the metadata from the video, rather than by directly searching for the keyword. MMToC parsed the content in lecture videos and generated an algorithm to index the topic in the video so that users can retrieve their interests conveniently and efficiently \cite{biswas2015mmtoc}. Aside from this, TalkMiner is a search engine for slide-based videos in which users can search for keywords from the video title and speech content or text in the slides \cite{adcock2010talkminer}. Based on the visual components of presentation slides, textual and mathematical phrases, speech contents, and mouse and cursor pointing motions acquired throughout the presentation, a visual interface for lecture video was proposed to extract various semantic clues for indexing video content and providing visual assistance \cite{zhao2019new}. There are several non-linear video search techniques. ViZig \cite{yadav2016vizig} identified the slide content and used anchor points to represent and visualize in the video timeline. Hierarchical visual interfaces have been proposed for drawing assistance \cite{dualface2021}, motion editing \cite{dualmotion21}, and document retrieval \cite{Matejka2021}. In this work, we aim to propose a hierarchical user interface for video retrieval and summarization.

\section{SYSTEM Overview}
In this work, we propose a hierarchical visual interface for lecture video retrieval and summarization. Before the proposal of user interface, we conducted a preliminary study about conventional video retrieval interfaces. 

\subsection{Preliminary Study}
We first conducted a preliminary investigation with a 7-Point Likert Scale questionnaire from 15 participants (college students around 20-years-old, 12 males and 3 females) with the following three questions (1 for strongly disagree and 7 for strongly agree). 

\noindent\textbf{Q 1.} Are you satisfied with the UI when you see the slide-based video on websites like YouTube?\\
\noindent\textbf{Q 2.} Do you think you are spending too much time searching for educational videos?\\
\noindent\textbf{Q 3.} If there is a better UI for retrieving and skimming educational videos, are you willing to try it?

\begin{figure} [ht]
   \begin{center}
   \begin{tabular}{c}
   \includegraphics[width=0.95\textwidth]{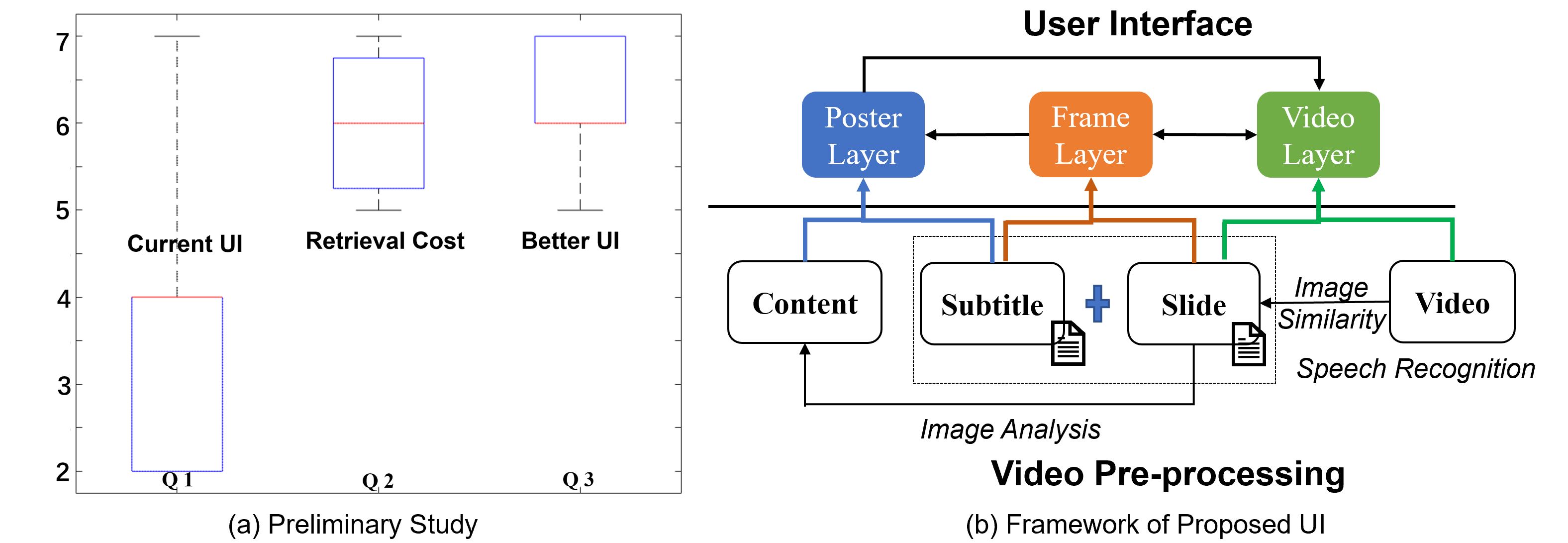}
   \end{tabular}
   \end{center}
   \caption[example] 
   { \label{fig:pre} 
Preliminary study about video retrieval interfaces (a) and the system framework of our proposed hierarchical user interface (b). }
 \end{figure}

We found that most participants felt that it is time consuming to search for educational videos were not satisfied with the current interface. As show in Fig.~\ref{fig:pre}(a), most participants are interested in trying a new interface if it can help them retrieve desired information more conveniently and efficiently. To solve these issues, we consider the hierarchical visual interface with three content layers. 

\subsection{System Framework}
In this work, the proposed architecture consists of three parts: the offline computation process for video pre-processing, construction of a video database, and a hierarchical user interface for educational videos as shown in Fig.~\ref{fig:pre}(b).
First, we extract slides, speech contents, and outlines from the educational video resource and store it into the database. The proposed interface consists of three layers: a video layer that contains video and groups of slides with timestamps, a frame layer that contains the speech content and visual content of one selected slide, and a poster layer that demonstrates the summarized video in a poster style.

When a user browses a video, the client will submit an HTTP request. The back-end then processes the request and returns JSON data to the client in the HTTP response. The front-end will then render the JSON data and generate a web page with the hierarchical visual interface.

\subsection{Video Pre-Processing}
\label{sec:4}
Video pre-processing plays a critical role in this system as it is a basic stage in listing and summarizing the video content show in Fig.~\ref{fig:pre}. Given an educational video, we first design an algorithm to compare the image hash of each frame in the video. If the differential of two frames is larger than the threshold, it means that the slide has changed; the slide will then be extracted and named with its timestamp in the video. During the extraction processing, we use the layoutparser \cite{shen2021layoutparser} function---a tool that comes with a set of layout data structures with carefully designed APIs that are optimized for document image analysis tasks: selecting layout/textual elements in the document, performing optical character recognition (OCR) for each detected layout region, visualizing the detected layouts, and loading layout data stored in JSON, CSV, and PDF---to extract titles, figures, and tables from the extracted slides. The layoutparser enables the extraction of complicated document structures using only several lines of code, with the help of state-of-the-art deep learning models. The extracted content is formed as a JSON file, in which "key" is the title and the "value" is the stored path for the other content. We thereafter download the subtitle of the corresponding video; if the video does not have subtitles, we use the state-of-the-art automatic speech recognition (ASR) approach\footnote{https://cloud.google.com/speech-to-text} to extract the speech contents. After that, we compare the timestamp between the speech content and the slide, and put the corresponding text into the JSON file. Upon completion of pre-processing, all JSON files are stored into the database and used to render the hierarchical visual interface. The title and its corresponding speech content, figures, and tables will render the poster layer page; the slide and subtitle will render the frame layer page and video layer page.
 
\subsection{Hierarchical User Interface}
\label{sec:5}
The hierarchical visual interface can help users retrieve a video conveniently and efficiently. When users search for a keyword, the system will retrieve the video title and its corresponding speech content from the database. The contents that meet the requirements will be rendered in the web page, as shown in Fig.~\ref{fig:ui}. When users browse a specific video, they will enter the hierarchical visual interface. The first page presented to the user is the poster layer.

In our survey of related work, we found that users are looking to use the textual content or the images in slides when they search for a specific topic in the video. Therefore, we design a poster layer that demonstrates the chapter of the video showing the figure in the slides and the speech content in the video in a poster format. This design can satisfy the users' browsing habits mentioned above. In case users want to see the detail of the point they are interested in, we bind the timestamp of the HTML data attribute of the title, figure, and speech content. When users click these elements (see red boxes in Fig.~\ref{fig:ui}), the website will redirect them to the corresponding time of the video in the video layer.

The left part of the video layer is a video player. The right part has a group of slides for the video, which the users can adjust as regards the number of slides shown. During the video playback, the slides on the right will change along with the video. To enable users to fast-forward the video to the part which they are interested in watching, we use the same design mentioned above. Thus, users can view the corresponding part in the video by clicking on the slide.

The frame layer fulfills the usage scenario, where users want to view the textual matter of the speech content or the slide rather than the video. Users can enter to the frame layer and view the textual matter of the speech content and slide by clicking the detail button in the slides. Users can also redirect to the other two layers by clicking on the corresponding buttons in the frame layer.

\section{User Study}

\begin{figure} [t]
   \begin{center}
   \begin{tabular}{c}
   \includegraphics[width=0.85\textwidth]{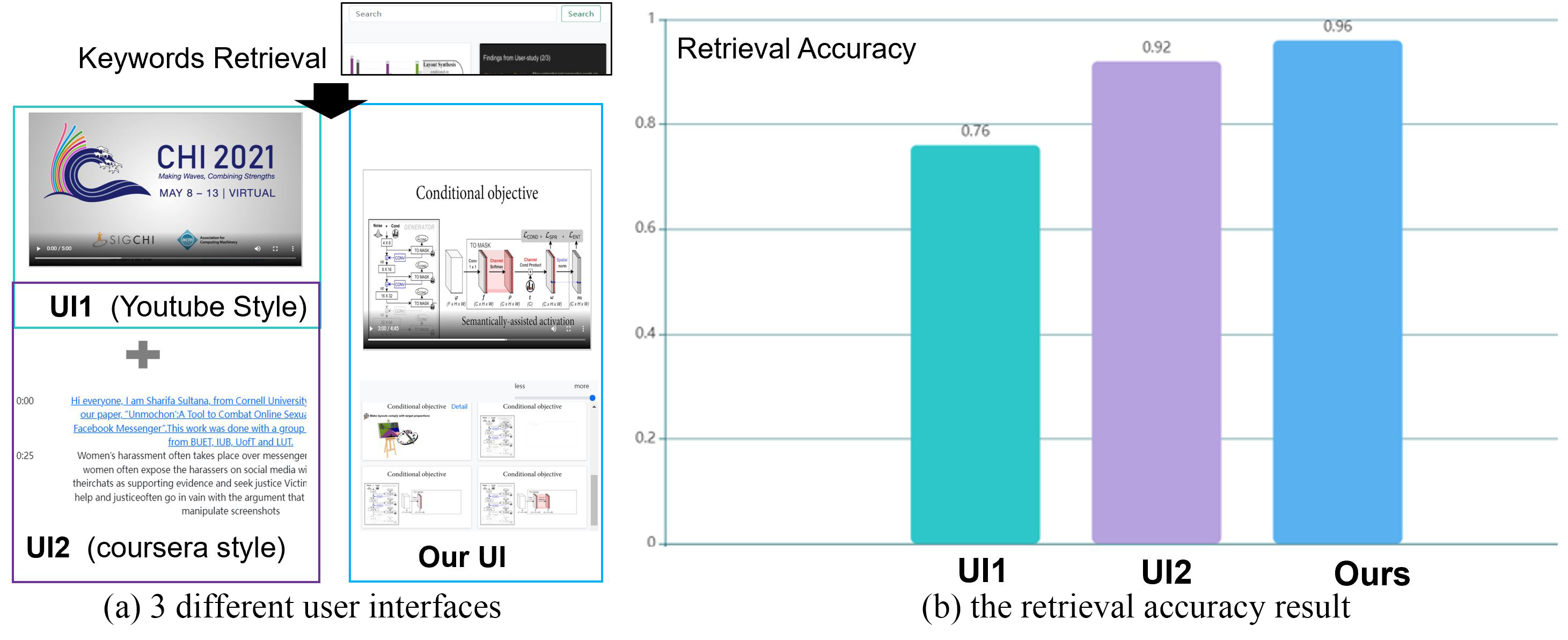}
   \end{tabular}
   \end{center}
   \caption[7] 
   { \label{fig:7} 
The result of comparison experiment. Top is the search result page. Left part is the Coursera style interface, and upper left is video only interface (a). The right figure shows the average accuracy for each interface (b). }
 \end{figure}

After the preliminary study, we conducted an experiment to compare the conventional retrieval interfaces and our interface. We invited 15 participants (college students around 20-years-old, 12 males and 3 females) in 3 groups. After a brief introduction of how to use our interface, we let each group use one of the following interfaces: YouTube style (video-only), Coursera style (see Fig.~\ref{fig:7}, video and subtitles), and our interface to retrieve the given five keywords in a limited time. For example, ``Find a video talking about `Image Synthesis'. '' All participants have not seen these videos before, and the duration of all the videos is around 5 minutes.

After the experiment, we asked them about their strategy in retrieving the video. We found that both our interface and Coursera style have indices which can help users locate the required information with more accuracy. However, as opposed to the Coursera style, our interface has more than one index, each of which can navigate to the other, making the retrieval processing more convenient. As demonstrated in Fig.~\ref{fig:7}, our interface achieved the highest accuracy results. The exact accuracy value is calculated by taking the average score from each interface.

\subsection{User Experience}
 In the second experiment, we conducted a user study to verify the user experiences by asking the participants to complete multiple video retrieval tasks. A general evaluation task consisting of visual search task, problem search task, and summarization task was used to test the participants. The visual search task requires participants to find a specific content in the video, for example, ``Find a slide where the author talks about the framework of this research. ''. The problem search task requires participants to answer a question relevant to the video, for example, ``What is the input in Partial-editing layer? ''. The summarization task requires participants to summarize the key points in the video. This time, we selected 8 participants from the previous experiment. After the experiment, they finished the questionnaire about their experience. The questionnaire uses a 7-Point Likert Scale ( 1 for strongly disagree and 7 for strongly agree).

\begin{table} [b]
   \begin{center}
   \begin{tabular}{c}
   \includegraphics[width=0.85\textwidth]{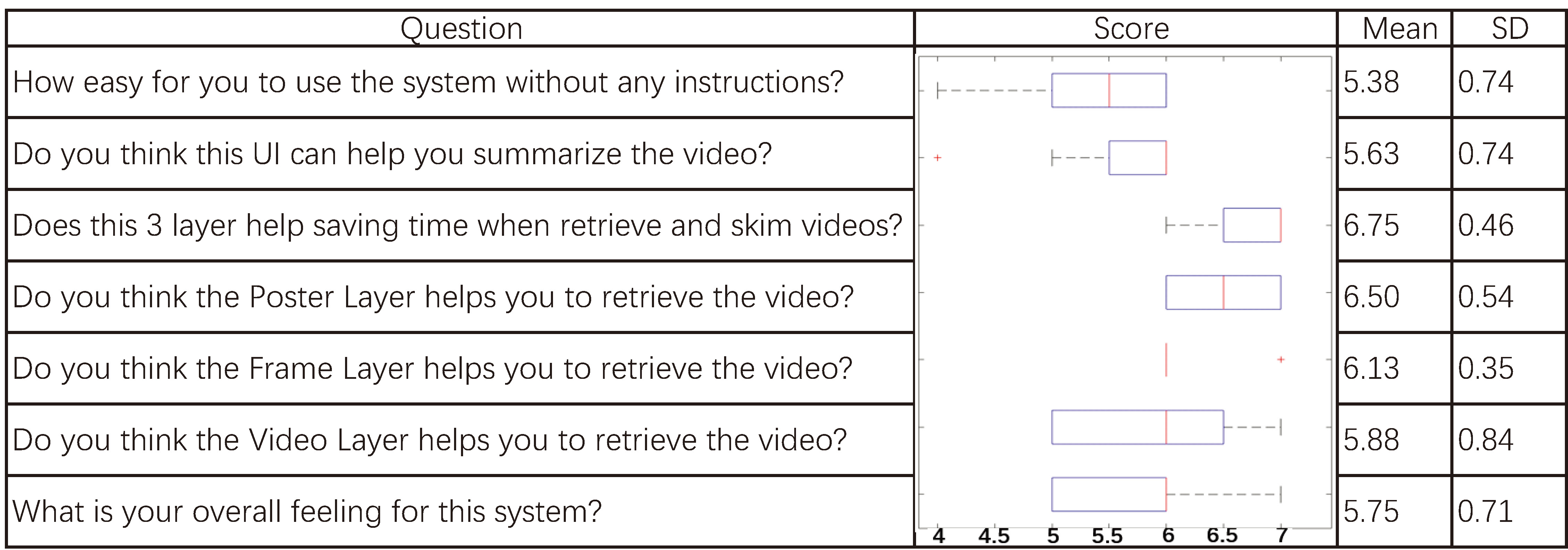}
   \end{tabular}
   \end{center}
   \caption[4] 
   { \label{fig:4} 
Results of the questionnaire.}
 \end{table}

Tab.~\ref{fig:4} shows the results of the experiment. The overview of our interface shows a better performance. As for the easy to use, participants are quite satisfied with our system. For the usefulness of the proposed 3-layer user interface, some participants felt the frame layer and poster layer can help them better retrieve the information than the video layer. In general, the results from our user study verified that the proposed interface can achieve high retrieval accuracy and good user experience.

\section{CONCLUSION}
This paper presented a hierarchical visual interface to retrieve and summarize educational lecture videos. We designed a three-layer user interface to fulfill users' searching intents during various usage scenarios. We then conducted a user study to verify our interface. In our user study, we demonstrated the usefulness of this three-layer interface for effective video navigation. It was shown to be more efficient than the outline-based video website offered by Massive Open Online Course (MOOC) websites and YouTube. Based on the feedback of our user study, this interface is anticipated to improve the video summarization through the poster layer in case of long duration videos (around one hour). It is also promising to improve the usability of the retrieval interface to fulfill users’ various retrieval intents in future work.

\section*{Acknowledgment}
We thank all participants in our user study. This work was supported by JAIST Research Grants, and JSPS KAKENHI Grant 20K19845, Japan.

% References
\bibliography{report} % bibliography data in report.bib
\bibliographystyle{spiebib} % makes bibtex use spiebib.bst

\end{document}